\pdfoutput=1

\documentclass[11pt]{article}

\usepackage{ACL2023}

\usepackage{bbm}
\usepackage{times}
\usepackage{cuted}
\usepackage{amsmath}
\usepackage{amssymb}
\usepackage{amsfonts}
\usepackage{latexsym}
\usepackage{booktabs}
\usepackage{multirow}
\usepackage{graphicx}

\usepackage[T1]{fontenc}

\usepackage[utf8]{inputenc}

\usepackage{microtype}
\usepackage{amsmath}

\usepackage{inconsolata}

\title{On the Limitations of Embedding Based Methods for Measuring Functional Correctness for Code Generation}

\author{Atharva Naik \\
  Carnegie Mellon University \\
  \texttt{arnaik@andrew.cmu.edu} \\
}

\begin{document}
\maketitle
\begin{abstract}
The task of code generation from natural language (NL2Code) has become extremely popular, especially with the advent of Large Language Models (LLMs).
However, efforts to quantify and track this progress have suffered due to a lack of reliable metrics for functional correctness.
While popular benchmarks like HumanEval have test cases to enable reliable evaluation of correctness, it is time-consuming and requires human effort to collect test cases.
As an alternative several reference-based evaluation metrics have been proposed, with embedding-based metrics like CodeBERTScore being touted as having a high correlation with human preferences and functional correctness.
In our work, we analyze the ability of embedding-based metrics like CodeBERTScore to measure functional correctness and other helpful constructs like editing effort by analyzing outputs of ten models over two popular code generation benchmarks. 
Our results show that while they have a weak correlation with functional correctness (0.16 $r_{bp}$), they are strongly correlated  (0.72$\tau$) with editing effort.
\end{abstract}

\section{Introduction}
Code generation sometimes also called program synthesis is the task of generating a program or code from a natural language (NL) intent or specification (NL2Code) \cite{code-intelligence-survey}.
Code generation has a lot of applications for both industry and academia due to the potential to increase the productivity of programmers \cite{sobania2022choose, productivity_barke}, connections with parsing \cite{Shin2019ProgramSA, sun2019grammar, ben2018neural}, machine reasoning \cite{gao2023pal}, planning \cite{singh2023progprompt}, mathematical capabilities \cite{Shao2024DeepSeekMathPT}, and creating structured representations for NLP tasks \cite{li2023codeie}.

Modeling approaches for code generation can be grouped into three categories \cite{code-intelligence-survey}: neural models of code, pre-trained models (CodePTMs), and Large Language Models for code (CodeLLMs). 
Neural methods included recurrent and convolutional networks possibly incorporating abstract syntax tree (AST) structure \cite{lmou_conv_ast, ast_neural}.
CodePTMs like CodeBERT \cite{feng-etal-2020-codebert}, CodeT5 \cite{Wang2021CodeT5IU}, and PLBART \cite{ahmad2021unified} were pre-trained with code-oriented self-supervised training objectives followed by task-specific fine-tuning.
Some CodePTMs also incorporate structures like AST, dataflow, and program dependency graphs (PDG) \cite{guo2020graphcodebert, guo2022unixcoder, wang2021syncobert}.
Among CodeLLMs, the earliest models were the Codex \cite{chen2021codex} and CodeGen \cite{nijkamp2022codegen} series. 
They were followed by BigCode project models like StarCoder \cite{li2023starcoder} and SantaCoder \cite{allalsantacoder} that adopted infilling (\cite{fried2022incoder}) in their training. 
Subsequently, CodeT5+ \cite{wang2023codet5+} and CodeLLaMA \cite{roziere2023code} became the first CodeLLMs to have an encoder-decoder architecture and to be trained from an existing model (LLaMA-2 \cite{touvron2023llama}) instead of from scratch respectively.
The DeepSeekCoder \cite{guo2024deepseek} model series added a focus on repository-level and cross-file code completion and the Lemur \cite{xu2023lemur} series also based on LLaMA-2 added a focus on agents with robust coding abilities and grounding in their environment.

Evaluation metrics in this space usually target the construct of functional correctness and can be categorized broadly into reference-based, reference-free, hybrid, and human evaluation. 
Reference-based methods like BLEU \cite{papineni-etal-2002-bleu}, ROUGE \cite{lin-2004-rouge}, METEOR \cite{banarjee2005} and chrF \cite{popovic-2015-chrf} utilize n-gram matching techniques.
Metrics like CodeBLEU \cite{ren2020codebleu}, RUBY \cite{tran2019does}, and CrystalBLEU \cite{crystalbleu} capture code-specific properties by adding dataflow, PDG, AST information or filtering out frequent n-grams.
However, prior work has shown that metrics like CodeBLEU and BLEU scores are not reliable indicators of functional correctness or human preferences \cite{chen2021codex, evtikhiev2023out}. 
Additionally, functionally non-equivalent programs can have higher BLEU scores than functionally equivalent ones \cite{chen2021codex}.
Embedding-based methods like BERTScore \cite{zhang2019bertscore}, COMET \cite{rei-etal-2020-comet}, and CodeBERTScore \cite{zhoucodebertscore} try to capture more semantics.
Reference-free methods include executing the generated code on test cases \cite{chen2021codex}, using LLMs with evaluation instructions like ICE-Score \cite{zhou2023language} and round trip evaluation \cite{allamanis2024unsupervised}.
An example of a hybrid method is CodeScore \cite{dong2023codescore} which trains a language model to capture execution semantics and can utilize both the NL context and reference code.
Finally despite higher costs human evaluation approaches like RealHumanEval \cite{mozannar2024realhumaneval} can capture more human-centered notions of productivity like time to complete coding tasks apart from functional correctness.

Despite the vast array of available approaches we choose to study embedding-based methods, like CodeBERTScore since among reference-based metrics they are the best at capturing semantic information, human preferences, and functional correctness \cite{zhoucodebertscore}, while being faster than execution-based evaluation and not requiring test-cases or test-case training data like CodeScore. 
They also require fewer computation resources like GPUs compared to LLM-based evaluation approaches like ICE-Score.
However, claims about the ability of embedding-based metrics to capture semantics and functional correctness warrant further investigation, since the underlying models (like CodeBERT) are known to fail in capturing semantic equivalence of code \cite{troshin-chirkova-2022-probing, Naik2022ProbingSG}.
In this work, we audit the ability of embedding-based metrics like CodeBERTScore to capture functional correctness for ten models over two popular code generation benchmarks.
We also analyze their effectiveness in measuring other more human-aligned constructs like editing effort \cite{dibia2023aligning}.
Our results show that embedding-based metrics indeed \textbf{fail to capture functional correctness} as measured by execution success, but can potentially function as metrics for editing effort. 
Our results motivate the importance of exploring more reference-free evaluation methods like CodeScore for code generation.
\section{Task Definition}
Before we begin our analysis it is important to formally define the task of code generation.
\\
\textbf{Context Space:} The context space $\mathcal{X}$ for code generation is the natural language instruction (NL) which the set of all strings can represent. $\mathcal{X} = \Sigma^* $
\\ 
\textbf{Decision Space:} The decision space $\mathcal{Y}$ for code generation is the space of all programs (PL) which can also be represented by the set of all strings, i.e. $\mathcal{Y} = \Sigma^*$
A more specific or constrained space can be defined for specific programming languages guided by the syntax and grammar \cite{ben2018neural, sun2019grammar, Shin2019ProgramSA}. 
Let's say the grammar $G$, with non-terminals $N$, a start symbol $S$, the set of terminals $\Sigma$, a finite set of production rules $P$ or $G = (N,\Sigma,P,S)$. We can define the decision space $\mathcal{Y}$ as the set of all possible programs that can be produced with the grammar or $\mathcal{Y} = \{w | S \Rightarrow_P^*w\}$
However, most state-of-the-art approaches like LLMs use the unconstrained decision space $\mathcal{Y} = \Sigma^*$.
\\
\textbf{Construct:} The main construct of interest is \textit{functional correctness}, i.e. whether the generated code captures the NL intent accurately.
It's important to measure because it can be challenging for humans to manually validate the correctness of generated code by eyeballing it which can lead to subtle bugs when using AI programming assistants \cite{Bar-Zik_2023}.
Since NL is an ambiguous and incomplete specification, practical efforts to measure functional correctness involve creating perfect specifications in the form of test cases on which generated code can be executed to verify the correctness, or reference programs that can be used as proxies to verify correctness.
Test case evaluation is more likely to be comprehensive (even though test suites can be limited \cite{liu2024your}) but is more expensive due to the human effort required to collect test cases, the time taken to execute code, and the need for sandboxing and environment setup to execute code \cite{yang2024intercode}.
Execution-based functional correctness is often measured with test cases using the $\text{pass@k}$ \cite{chen2021codex} metric which is defined as:
\[ \text{pass@k} := \mathbb{E}_{\text{problems}}\left[1-\frac{\binom{n-c}{k}}{\binom{n}{k}}\right] \]
Here $k$ is the number of programs sampled from the model, while $c$ is the number of correct programs that pass all test cases.
A high pass rate means a high probability that the generated code is functionally correct.

Another interesting construct is the \textit{editing effort} (or syntactic similarity) highlighted by \cite{dibia2023aligning}.
They show that functional correctness can underestimate productivity gains as outputs/system decisions that fail test cases can still be useful if the overall effort for coding can be reduced.
They propose $\text{EDIT-SIM}$ or the normalized edit similarity defined below as a proxy for effort:
\[
\text{EDIT-SIM} = 1 - \frac{lev(gen,ref)}{max(len(gen),len(ref))}
\]
where $gen$ is the code generated or system decision, $ref$ is a reference solution to the problem, and $lev$ is the character Levenshtein edit distance \cite{Levenshtein_SPD66}.
A high $EDIT-SIM$ is more desirable and means low editing effort.

Similar to functional correctness, \textit{distinguishability} (d) \cite{crystalbleu} measures the ability of a metric to detect the similarity of semantically similar code over semantically different code. 
It can be mathematically formulated as:
\[
d = \frac{m(Pairs_{intra})}{m(Pairs_{inter})}
\]
Where $m$ is the metric, where $Pairs_{intra}$ represents the intra-class similarity between pairs of semantically similar code (belonging to an equivalence class or partition based on functional properties) while $Pairs_{inter}$ represents the similarity between pairs of code belonging to different equivalence classes.
A high distinguishability is ideal as it means the metric can cluster semantically similar codes together.

However \cite{zhoucodebertscore} shows that since distinguishability uses absolute metric values it can be easily gamed by simply exponentiating metric values making it a bad proxy or alternative for functional correctness.
Additionally, they speculate that any meta-metric or construct of correctness that compares exact scores is likely to be gameable compared to ranking-based approaches.
Hence we don't use it for our analysis in this study.

\textbf{Metric of study:}
The CodeBERTScore metric proposed by \cite{zhoucodebertscore} as a code-specific variant of BERTScore \cite{zhang2019bertscore} utilizes embeddings of a pre-trained CodeBERT model \cite{feng-etal-2020-codebert} to compute the similarity of a reference code ($y^*$) with a candidate ($\hat{y}$).
They compute all pairs of token similarity between the candidate and reference while masking out certain tokens (according to masks $m^*$ and $\hat{m}$) like punctuation to compute the precision (P) (matching candidate tokens $\hat{y}_j$ to most similar reference tokens $y^*_i$) (Eq. \ref{eq:P}) and recall (R) (Eq. \ref{eq:R}) (matching reference tokens to most similar candidate tokens). 
The F1 (Eq. \ref{eq:F1}) and F3 (Eq. \ref{eq:F3}) scores are computed using the precision and recall scores with the F3 score weighing the recall more heavily.

\begin{strip}
\begin{equation}
\text{CodeBERTScore}_P = \frac{1}{|\hat{y}[\hat{\mathbf{m}}]|} \sum_{\hat{y}_j \in \hat{y}[\hat{\mathbf{m}}]} \max_{y_i^* \in y^*[\mathbf{m^*}]} sim(y_i^*,\hat{y}_j)
\label{eq:P} 
\end{equation}
\begin{equation}
\text{CodeBERTScore}_R = \frac{1}{|y^*[\mathbf{m}]|} \sum_{y_i^* \in y^*[\mathbf{m}^*]} \max_{\hat{y}_j \in \hat{y}[\hat{\mathbf{m}}]} sim(y_i^*,\hat{y}_j)   
\label{eq:R} 
\end{equation}
\begin{equation}
\text{CodeBERTScore}_{F_1} = \frac{2 \cdot \text{CodeBERTScore}_P \cdot \text{CodeBERTScore}_R}{\text{CodeBERTScore}_P + \text{CodeBERTScore}_R}
\label{eq:F1} 
\end{equation}
\begin{equation}
\text{CodeBERTScore}_{F_3} = \frac{10 \cdot \text{CodeBERTScore}_P \cdot \text{CodeBERTScore}_R}{9 \cdot \text{CodeBERTScore}_P + \text{CodeBERTScore}_R}
\label{eq:F3}  
\end{equation}
\end{strip}
\section{Related Work}


\subsection{Datasets}
Due to the popularity of code generation as a task, many datasets exist out there. 
Early datasets like CoNaLa \cite{yin2018mining} and its multilingual version mCoNaLa \cite{wang2023mconala} targeted the task of generating small code snippets.
This was followed by the development of some of the most popular function-generation benchmarks like APPS \cite{hendrycks2021measuring}, MBPP \cite{austin2021program}, and HumanEval \cite{chen2021codex} with problems that span basic ``closed-domain'' algorithmic challenges in Python with test cases to support evaluation.
HumanEval+ is an extended version of the HumanEval dataset which has better test case coverage \cite{liu2024your}, while HumanEval-X \cite{zheng2023codegeex} and MultiPL-E \cite{cassano2022multipl} (extends HumanEval and MBPP) contain additional programming languages other than Python.
Code generation datasets for more complex contexts like Jupyter notebooks (ARCADE \cite{yin2023natural}, ExeDS \cite{huang-etal-2022-execution} and JuICe \cite{agashe-etal-2019-juice}) and classes (CoderEval \cite{yu2023codereval} and ClassEval \cite{du2023classeval}) have been proposed to move beyond function level code generation. 
Other datasets aim for specialized domains like data-science DS-1000 \cite{lai2023ds} or open domain code generation \cite{wang2022execution}.
Some recent datasets have also focused on interactive and multi-turn coding datasets like InterCode \cite{yang2024intercode} and CodeClarQA \cite{li-etal-2023-python} (clarification question generation).
Finally, some of the latest code generation datasets target repository level and code agent tasks like SWEBench \cite{Jimenez2023SWEbenchCL}  RepoEval \cite{zhang-etal-2023-repocoder}, CodeAgentBench \cite{zhang2024codeagent}, RepoBench \cite{liu2023repobench} and Stack-Repo \cite{shrivastava2023repofusion}.

\subsection{Metrics}



For this study, we mainly focus on reference-based evaluation metrics, as the embedding-based methods belong to that category. Each metric is described below:
\\
\textbf{Exact match:}
\[ \text{EM} := \mathbb{E}_{\text{problems}}\left[\mathbbm{1}_{cand = ref}\right] \]
Here the $\mathbbm{1}_{c \in \mathcal{T}}$ is an indicator variable which is $1$ when the candidate is the same as the reference. 
This essentially plays the role of a lower limit on the performance as in every instance where the exact match is $1$ all other metrics will also attain their highest value.
\\
\textbf{BLEU score:}
\[ \text{BLEU} = \text{BP} \times \exp\left(\sum_{n=1}^{N} w_n \cdot \log \left(\frac{\text{C}_{\text{clip}, n}}{\text{C}_{\text{total}, n}}\right)\right) \]
\\
Where: \\
- \(\text{BP}\) is the Brevity Penalty \\
- \(N\) is the maximum n-gram order considered (typically \(N=4\)) \\
- \(w_n\) is the weight assigned to the \(n\)-gram precision \\
- \(\text{C}_{\text{clip}, n}\) is the clipped count of \(n\)-grams in the generated code \\
- \(\text{C}_{\text{total}, n}\) is the total count of \(n\)-grams in the generated code.
\\
The Brevity Penalty (\(\text{BP}\)) is defined as:
\\
\[ \text{BP} =
\begin{cases} 
1 & \text{if } \text{len}_{\text{gen}} > \text{len}_{\text{ref}} \\
e^{(1 - \frac{\text{len}_{\text{ref}}}{\text{len}_{\text{gen}}})} & \text{if } \text{len}_{\text{gen}} \leq \text{len}_{\text{ref}}
\end{cases}
\]
\\
\textbf{CrystalBLEU Score:}
\[ \text{CrystalBLEU} = \text{BP} \times \exp\left(\sum_{n=1}^{N} w_n \cdot \log \left(\frac{\text{C}_{\text{clip}, S_k, n}}{\text{C}_{\text{total},S_k, n}}\right)\right) \]
The Crystal BLEU metric is defined almost identically to BLEU but with one major difference in the clipped n-gram precision computation.
The modified precision now also takes in $S_k$ or the top-k most frequent n-grams ($k=50$ for our study) and removes them from the n-gram counts from both the numerator and denominator in the precision.
\\
\textbf{CodeBLEU:} 
\[ \text{CodeBLEU} = \alpha \cdot \text{BLEU} + \beta \cdot \text{BLEU}_{\text{weight}} + \gamma \cdot \text{Match}_{\text{ast}} + \delta \cdot \text{Match}_{\text{df}} \]    
The CodeBELU score adds a weighted BLEU score, syntax match score, and semantic dataflow match score to the BLEU score.

Here $\text{BLEU}_{\text{weight}}$ is the weighted n-gram BELU score:

\[ \text{BLEU}_{\text{weight}} = \text{BP} \cdot \exp\left(\sum_{n=1}^Nw_n\log p_n\right) \]

Here $p_n$ is the weighted n-gram precision which is defined as: 
\[ p_n = \frac{\sum_{C \in \text{Candidates}}\sum_{i=1}^l \mu_n^i \cdot \text{Count}_{\text{clip}}(C(i,i+n))}{\sum_{C' \in \text{Candidates}}\sum_{i=1}^l \mu_n^i \cdot \text{Count}_{\text{clip}}(C'(i,i+n))} \]

Where: \\
- $C(i,i+n)$  is the n-gram from the position $i$ to the position $i + n$ \\
- $\text{Count}_{\text{clip}}(C(i,i+n))$ is the maximum number of n-grams co-occurring in a candidate code and a set of reference codes. \\
- $\mu^i_n$ denotes the weights of different keywords or n-grams. The weight of the keywords is usually set to 5 times the weight of other tokens. \\
- $\text{BP}$ is the brevity penalty as defined for BLEU. 

The syntactic match score $\text{Match}_{\text{ast}}$ is defined below:

\[ \text{Match}_{\text{ast}} = \text{Count}_{\text{clip}}(T_{\text{cand}})/\text{Count}(T_{\text{ref}}) \]
Where: \\
- $\text{Count}(T_{\text{ref}})$ is the total number of the reference subtrees. \\
- $\text{Count}_{\text{clip}}(T_{\text{cand}})$ is the number of the candidate subtrees that are matched to the reference. \\ 

Finally, the data match score $\text{Match}_{\text{df}}$ is defined below:    

\[ \text{Match}_{\text{df}} = \text{Count}_{\text{clip}}(DF_{\text{cand}})/\text{Count}(DF_{\text{ref}}) \]
Where: \\
- $\text{Count}(DF_{\text{df}})$ is the total number of the reference data-flows. \\
- $\text{Count}_{\text{clip}}(DF_{\text{cand}})$ is the number of matched candidate data-flows. 
\\ 
\textbf{chrF:}
\begin{strip}
    \begin{align*}
\text{CHRF} = \max \left( 1 - \frac{2 \cdot \text{unigram\_errors}}{\text{reference\_length} + \text{hypothsis\_length}}, 0 \right) \cdot \left( 1 - \frac{\text{abs}(r - h)}{r + h} \right)^\beta
    \end{align*}
\end{strip}

\textbf{BERTScore:}
The BERTScore metric is computed similarly to CodeBERTScore and we use the F1 measure (Eq. \ref{eq:F1}) throughout the study.
\section{Methods}
To measure the capacity of code generation metrics to capture functional correctness we use the datasets and systems described below:

\subsection{Datasets}
We choose the HumanEval \cite{chen2021codex} and MBBP \cite{austin2021program} benchmarks because of their popularity in the research community as standard benchmarks, relatively simpler coding tasks that don't involve external libraries, and the availability of test cases for directly measuring functional correctness.
The HumanEval dataset has 164 examples and an average of 7.7 test cases.
The MBPP (sanitized) test set has 257 examples with 3 test case on average per instance.

\subsection{Systems}
To analyze the effectiveness of the evaluation metrics we pick some of the most popular open-source language models (StarCoder2 \cite{lozhkov2024starcoder}, CodeLLaMA \cite{roziere2023code}, WizardCoder \cite{luo2023wizardcoder}, DeepSeekCoder \cite{guo2024deepseek}, Magicoder \cite{wei2023magicoder}, CodeQwen \cite{qwen} CodeGemma \cite{team2024gemma}, LLaMA-3 \cite{llama3modelcard}) in the 6.7B-15B parameter range as well as some popular closed source models (GPT-3.5-Turbo \cite{gpt3.5} and GPT-4 \cite{gpt4}). 
The results obtained by the models on all the metrics and the constructs of interest (functional correctness and editing effort) for HumanEval and MBPP are shown in Table~\ref{tab:all_humaneval_results} and Table~\ref{tab:all_mbpp_results} respectively. 
The CodeBERTScore metrics like CodeBERTScore-P etc. are abbreviated as CB-P and so on.

\begin{table*}[]
\centering
\resizebox{\textwidth}{!}{
\begin{tabular}{@{}lrrrrrrrrrrrr@{}}
\toprule
\multirow{2}{*}{\textbf{Model}} & \multicolumn{10}{c}{\textbf{Metrics}}                                         & \multicolumn{2}{c}{\textbf{Constructs}} \\ \cmidrule(l){2-13} 
 &
  EM &
  chrF &
  BLEU &
  CodeBLEU &
  CrystalBLEU &
  BERTScore &
  CB-P &
  CB-R &
  CB-F1 &
  CB-F3 &
  \begin{tabular}[c]{@{}r@{}}Functional \\ Correctness\end{tabular} &
  \begin{tabular}[c]{@{}r@{}}Editing \\ Effort\end{tabular} \\ \midrule
StarChat2-15B                   & 0.000 & 0.238 & 0.111 & 0.188 & 0.030 & 0.570 & 0.479 & 0.525 & 0.500 & 0.519 & 0.366              & 0.191              \\
CodeLLaMA-13B-Hf                & 0.000 & 0.156 & 0.032 & 0.063 & 0.008 & 0.835 & 0.743 & 0.673 & 0.705 & 0.679 & 0.018              & 0.221              \\
WizardCoder-13B                 & 0.000 & 0.191 & 0.058 & 0.105 & 0.029 & 0.858 & 0.783 & 0.700 & 0.737 & 0.707 & 0.024              & 0.255              \\
Magicoder-S-DS-6.7B             & 0.030 & 0.398 & 0.251 & 0.248 & 0.070 & 0.897 & 0.816 & 0.803 & 0.809 & 0.804 & 0.543              & 0.440              \\
CodeGemma-7B-It                 & 0.018 & 0.381 & 0.213 & 0.287 & 0.054 & 0.893 & 0.774 & 0.832 & 0.799 & 0.824 & 0.500              & 0.369              \\
DeepSeekCoder-6.7B-Instruct &
  0.030 &
  0.427 &
  0.260 &
  0.267 &
  0.084 &
  \textbf{0.906} &
  \textbf{0.838} &
  0.829 &
  \textbf{0.832} &
  0.829 &
  0.598 &
  \textbf{0.459} \\
LLaMA-3-8B-Instruct             & 0.012 & 0.368 & 0.199 & 0.269 & 0.053 & 0.890 & 0.785 & 0.821 & 0.800 & 0.816 & 0.488              & 0.377              \\
CodeQwen1.5-7B-Chat &
  \textbf{0.043} &
  \textbf{0.433} &
  \textbf{0.267} &
  \textbf{0.311} &
  \textbf{0.101} &
  0.904 &
  0.820 &
  \textbf{0.846} &
  0.830 &
  \textbf{0.843} &
  \textbf{0.652} &
  \textbf{0.459} \\ \midrule
GPT-3.5-Turbo                   & 0.024 & 0.433 & 0.266 & 0.284 & 0.080 & 0.904 & 0.840 & 0.836 & 0.838 & 0.837 & 0.567              & 0.456              \\
GPT-4 &
  \textbf{0.055} &
  \textbf{0.482} &
  \textbf{0.325} &
  \textbf{0.323} &
  \textbf{0.107} &
  \textbf{0.914} &
  \textbf{0.861} &
  \textbf{0.855} &
  \textbf{0.857} &
  \textbf{0.855} &
  \textbf{0.665} &
  \textbf{0.499} \\ \bottomrule
\end{tabular}
}
\caption{Results for all models over all metrics and constructs for the HumanEval dataset. The CodeQwen and DeepSeekCoder seem to have the best performance over the embedding-based metrics and n-gram-based metrics respectively and the functional correctness and editing effort constructs respectively. Among the close source models, GPT-4 is better on all metrics and constructs and better overall among all models.}
\label{tab:all_humaneval_results}
\end{table*}

\begin{table*}[]
\centering
\resizebox{\textwidth}{!}{
\begin{tabular}{@{}lrrrrrrrrrrrr@{}}
\toprule
\multirow{2}{*}{\textbf{Model}} & \multicolumn{10}{c}{\textbf{Metrics}}                                         & \multicolumn{2}{c}{\textbf{Constructs}} \\ \cmidrule(l){2-13} 
 &
  EM &
  chrF &
  BLEU &
  CodeBLEU &
  CrystalBLEU &
  BERTScore &
  CB-P &
  CB-R &
  CB-F1 &
  CB-F3 &
  \begin{tabular}[c]{@{}r@{}}Functional \\ Correctness\end{tabular} &
  \begin{tabular}[c]{@{}r@{}}Editing \\ Effort\end{tabular} \\ \midrule
StarChat2-15B                   & 0.000 & 0.353 & 0.151 & 0.250 & 0.016 & 0.893 & 0.798 & 0.820 & 0.807 & 0.817 & 0.533              & 0.353              \\
CodeLLaMA-13B-Hf                & 0.000 & 0.201 & 0.032 & 0.044 & 0.000 & 0.868 & 0.806 & 0.712 & 0.755 & 0.720 & 0.027              & 0.239              \\
WizardCoder-13B                 & 0.000 & 0.207 & 0.036 & 0.043 & 0.000 & 0.861 & 0.802 & 0.709 & 0.752 & 0.717 & 0.012              & 0.237              \\
Magicoder-S-DS-6.7B             & 0.000 & 0.360 & 0.160 & 0.224 & 0.018 & 0.898 & 0.838 & 0.815 & 0.825 & 0.817 & 0.529              & 0.383              \\
CodeGemma-7B-It                 & 0.000 & 0.299 & 0.085 & 0.241 & 0.009 & 0.861 & 0.718 & 0.785 & 0.749 & 0.778 & 0.494              & 0.229              \\
DeepSeekCoder-6.7B-Instruct     & 0.000 & 0.317 & 0.128 & 0.180 & 0.008 & 0.889 & 0.814 & 0.792 & 0.802 & 0.794 & 0.366              & 0.352              \\
LLaMA-3-8B-Instruct             & 0.000 & 0.349 & 0.140 & 0.221 & 0.015 & 0.887 & 0.810 & 0.810 & 0.809 & 0.809 & 0.451              & 0.357              \\
CodeQwen1.5-7B-Chat &
  \textbf{0.004} &
  \textbf{0.487} &
  \textbf{0.317} &
  \textbf{0.306} &
  \textbf{0.066} &
  \textbf{0.910} &
  \textbf{0.854} &
  \textbf{0.863} &
  \textbf{0.857} &
  \textbf{0.862} &
  \textbf{0.619} &
  \textbf{0.474} \\ \midrule
GPT-3.5-Turbo                   & 0.000 & 0.445 & 0.246 & 0.249 & 0.025 & 0.904 & 0.845 & 0.845 & 0.844 & 0.845 & 0.786              & 0.426              \\
GPT-4 &
  \textbf{0.004} &
  0.469 &
  0.301 &
  0.267 &
  0.051 &
  \textbf{0.911} &
  \textbf{0.873} &
  0.854 &
  \textbf{0.862} &
  0.856 &
  \textbf{0.802} &
  0.458 \\ \bottomrule
\end{tabular}
}
\caption{Results for all models over all metrics and constructs for the MBPP dataset. The CodeQwen model has the best performance over all metrics among the open-source models and the constructs. Among the close source models, GPT-4 is better on all metrics and best for most metrics overall (except the BLEU metrics, chrF and CodeBERTScore-F3).}
\label{tab:all_mbpp_results}
\end{table*}
\section{Metric Properties}
We analyze and compare various properties of CodeBERTScore with respect to other metrics.
\subsection{Boundedness}
The CodeBERTScore metric theoretically spans values from $-1$ to $1$ similar to BERTScore \cite{githubBert_scorejournalrescale_baselinemdMaster}, but empirically it only spans from $0$ to $1$ empirically. The bounds for all metrics are shown in Table~\ref{tab:metric_bounds}.

\begin{table*}[]
\centering
\begin{tabular}{@{}lrrrr@{}}
\toprule
\multirow{2}{*}{\textbf{Metric}} & \multicolumn{2}{c}{\textbf{Theoretical Bounds}} & \multicolumn{2}{c}{\textbf{Empirical Bounds}} \\ \cmidrule(l){2-5} 
                                 & \textbf{Max}           & \textbf{Min}           & \textbf{Max}          & \textbf{Min}          \\ \midrule
Exact Match      & 0  & 1 & 0 & 1 \\
chrF             & 0  & 1 & 0 & 1 \\
BLEU             & 0  & 1 & 0 & 1 \\
CodeBLEU         & 0  & 1 & 0 & 1 \\
CrystalBLEU      & 0  & 1 & 0 & 1 \\
BERTScore-F1     & -1 & 1 & 0 & 1 \\
CodeBERTScore-P  & -1 & 1 & 0 & 1 \\
CodeBERTScore-R  & -1 & 1 & 0 & 1 \\
CodeBERTScore-F1 & -1 & 1 & 0 & 1 \\
CodeBERTScore-F3 & -1 & 1 & 0 & 1 \\ \bottomrule
\end{tabular}
\caption{The theoretical bounds and empirically attained bounds for all metrics}
\label{tab:metric_bounds}
\end{table*}

\subsection{Score Distribution}
\begin{figure*}[!tbh]
    \centering
    \includegraphics[width=\textwidth]{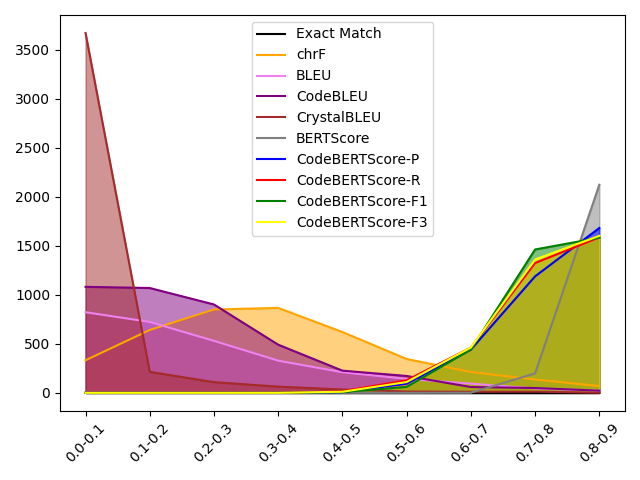}
    \caption{The distribution of metric values. The \textbf{x-axis} shows the range of values while the \textbf{y-axis} shows the number of model decisions with values in that range. The distributions clearly show that the embedding metrics tend to over-reward decisions while the other metrics tend to over-penalize them. The \textbf{chrF} metric has the least skew}
    \label{fig:score_dist}
\end{figure*}

We also analyze the score distribution of all the metrics in Figure~\ref{fig:score_dist} and summarize the centrality and shape measures in Table~\ref{tab:score_dist}. 
The score distributions show that the embedding-based metrics BERTScore and CodeBERTScore overly reward models while the other metrics, especially exact match overly penalize the models.  
The centrality measures like mean, median, and mid-hinge also point towards that.
The Kurtosis (excess) measures for all metrics except chrF are Leptokurtic, meaning they have fatter tails than a normal distribution.
The chrF metric exhibits the least skewness and excess Kurtosis while the exact match metric is most skewed with a fat tail, as most of the values are zero (high peak) and then there is a fait tail of all the non-zero values.

\begin{table*}[]
\centering
\begin{tabular}{@{}lrrrrrr@{}}
\toprule
\textbf{Metric} & \textbf{Median} & \textbf{Midhinge} & \textbf{Mean} & \textbf{Std. Dev.} & \textbf{Skewness} & \textbf{Kurtosis} \\ \midrule
Exact Match      & 0     & 0     & 0.009 & 0.093 & 10.526 & 108.793 \\
chrF             & 0.323 & 0.330 & 0.350 & 0.205 & 0.813  & 0.645   \\
BLEU             & 0.108 & 0.132 & 0.175 & 0.212 & 1.611  & 2.642   \\
CodeBLEU         & 0.186 & 0.189 & 0.215 & 0.177 & 1.314  & 2.302   \\
CrystalBLEU      & 0 & 0 & 0.037 & 0.117 & 4.772  & 27.645  \\
BERTScore-F1     & 0.892 & 0.889 & 0.876 & 0.118 & -6.035 & 42.250  \\
CodeBERTScore-P  & 0.817 & 0.815 & 0.800 & 0.135 & -3.072 & 15.600  \\
CodeBERTScore-R  & 0.806 & 0.804 & 0.789 & 0.137 & -2.829 & 13.819  \\
CodeBERTScore-F1 & 0.804 & 0.804 & 0.792 & 0.131 & -3.187 & 17.127  \\
CodeBERTScore-F3 & 0.805 & 0.803 & 0.790 & 0.135 & -2.937 & 14.796  \\ \bottomrule
\end{tabular}
\caption{The centrality and shape measures for all the metrics. The metrics highlighted in red are heavily skewed towards the left (i.e. they over-penalize the models) while the ones highlighted in red are heavily skewed towards the right (i.e. they over-reward the models). Additionally almost all of the metrics are Leptokurtic, i.e. they have fatter tails than a normal distribution, especially Exact Match and BERT-Score.}
\label{tab:score_dist}
\end{table*}

\subsection{Number of ties}
We compute the percentage of ties across a given context for all pairs of model comparisons and show the results in Table~\ref{tab:number_of_ties}. 
For continuous metrics that can attain any real values between zero to one on an instance level, we count it as a tie if the metric values differ by a small value $\epsilon = 10^{-6}$.
The exact match metric has the most ties because of sparsity and being $0$ for most cases while, BERTScore and CodeBERTScore metrics have fewest ties.

\begin{table}[]
\centering
\begin{tabular}{@{}lr@{}}
\toprule
\textbf{Metric}  & \multicolumn{1}{c}{\textbf{Rate of Ties}} \\ \midrule
Exact Match      & 98.8                                      \\
chrF             & 3.73                                      \\
BLEU             & 16.84                                     \\
CodeBLEU         & 7.7                                       \\
CrystalBLEU      & 47.77                                     \\
BERTScore-F1     & 3.17                                      \\
CodeBERTScore-P  & 3.17                                      \\
CodeBERTScore-R  & 3.17                                      \\
CodeBERTScore-F1 & 3.18                                      \\
CodeBERTScore-F3 & 3.17                                      \\ \bottomrule
\end{tabular}
\caption{Percentage of ties for each metric. For real-valued metrics, we count it as a tie if the metric values differ by a very small value $\epsilon = 10^{-6}$. The exact match metric has the most ties because of sparsity and being $0$ for most cases while BERTScore and CodeBERTScore metrics have the fewest ties.}
\label{tab:number_of_ties}
\end{table}

\subsection{Convergent and Discriminant Validity}
\begin{figure}[!tbh]
    \centering
    \includegraphics[width=0.5\textwidth]{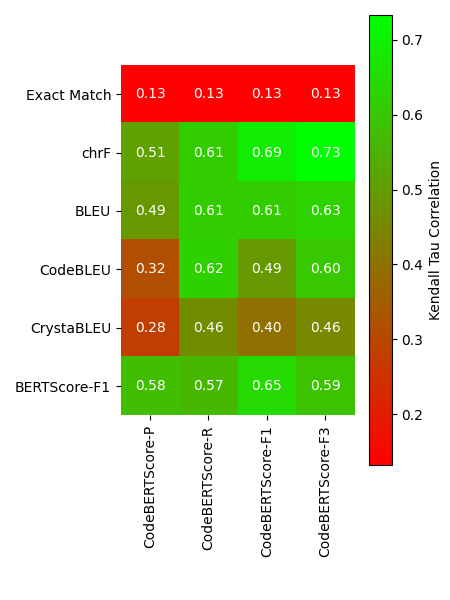}
    \caption{Correlations of the CodeBERTScore metrics with the other metrics. Except for exact match, most of the metrics have a moderate correlation with CodeBERTScore}
    \label{fig:conv_validity_heatmap}
\end{figure}

We analyze the Kendall Tau ($\tau$) correlation between all the metrics (EM, chrF, BLEU, CodeBLEU, CrystalBLEU, and BERTScore-F1) and all the CodeBERTScore metrics (P, R, F1, F3) and visualize the results via a heatmap in Figure~\ref{fig:conv_validity_heatmap}.
All the correlations are statistically significant with very low p-values.
We observe that except for exact match, CodeBERTScore exhibits moderate correlation with all metrics.

\subsection{Discriminative Power}
\begin{figure*}[!tbh]
    \centering
    \includegraphics[width=\textwidth]{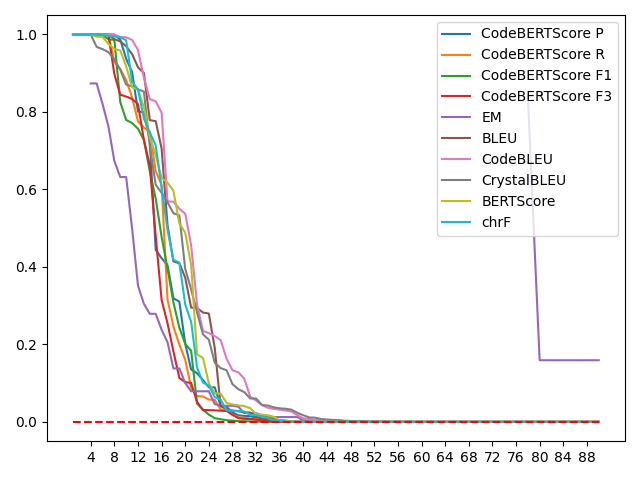}
    \caption{The achieved significance level of the metric values. \textbf{x-axis} shows pairs of runs while \textbf{y-axis} shows the p-values. The dotted line shows the acceptance threshold}
    \label{fig:asl_unclipped}
\end{figure*}

\begin{figure*}[!tbh]
    \centering
    \includegraphics[width=\textwidth]{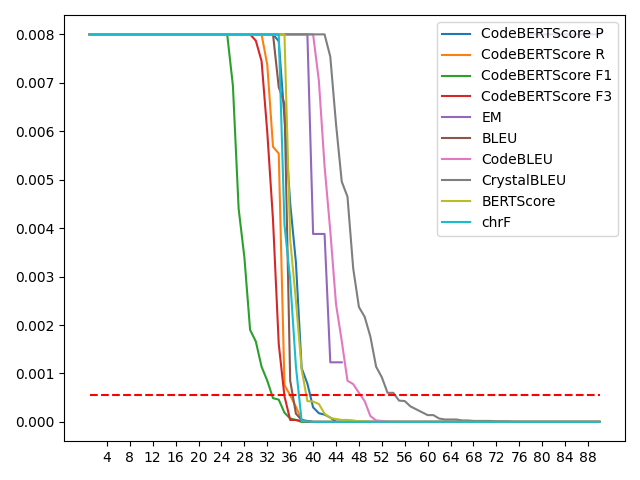}
    \caption{The achieved significance level of the metric values. \textbf{x-axis} shows pairs of runs while \textbf{y-axis} shows the p-values. The dotted line shows the acceptance threshold. All p-values above $0.008$ are clipped to zoom into the portion close to the acceptance threshold.}
    \label{fig:asl_clipped}
\end{figure*}

We plot the discriminative power of all the metrics compared using the student t-test with the hypothesis that the mean of the underlying distribution of the first model is less than the mean of the distribution of the second model with Bonferroni correction \cite{Armstrong2014-nh} applied to the acceptance threshold $\alpha = \frac{0.05}{n}$ ($n$ is the number of hypotheses) to account for multiple hypotheses.
Since there are $n=2\times{10 \choose 2}=90$ comparisons, the acceptance threshold $\alpha = \frac{0.05}{90} = 0.00055$.
We plot the achieved significance level \cite{sakai2014metrics} for each metric (sorted p-values on the y-axis and run comparison pairs on the x-axis) in Figure~\ref{fig:asl_unclipped}.
However, since the acceptance threshold is very low, to make the plots easier to visualize we also show a clipped version where all the p-values are clipped if they are above $0.008$ in Figure~\ref{fig:asl_clipped}.
We note that CodeBERTScore-F1 has the most discriminative power being able to achieve statistical significance for 58/90 pairs, and the exact match being the worst, not being able to distinguish any pairs.
Additionally, the code-specific BLEU scores (CodeBLEU, CrystalBLEU) are worse than regular BLEU which is comparable to BERTScore and chrF.

\section{Empirical Analysis}
\subsection{Hypotheses}
Since the primary focus of the work is to test the reliability of CodeBERTScore as a measure of the primary construct of functional correctness and the secondary measure of editing effort, we can formulate the following hypotheses: \\
\textbf{H1.} CodeBERTScore is moderate to strongly correlated with execution results ($\text{pass@1}$).
\\
\textbf{H2.} CodeBERTScore is robust to surface-level perturbation, i.e. it exhibits strong auto-correlation under function-name and variable input perturbations (test of criterion validity with invariant transformations)
\\
\textbf{H3.} CodeBERTScore is moderate to strongly correlated with editing effort ($\text{EDIT-SIM}$)

\subsection{Hypothesis Testing}
To test \textbf{H1} we measure the correlation between the $\text{pass@1}$ metric and CodeBERTScore-F1 using point-biserial correlation $r_{bp}$ which is suitable for measuring the correlation between a binary or dichotomous variable like whether an instance passes all test cases and a continuous variable like CodeBERTScore. 
We use F1 for all subsequent hypotheses as it is the most sensitive metric according to the discriminative power and combines both precision and recall without any specific bias to either.
In addition to this correlation, we also compute the point-biserial correlation between editing effort ($\text{EDIT-SIM}$) and $\text{pass@1}$ and BERTScore and $\text{pass@1}$ to see if the two constructs are correlated and compare the reliability of BERTScore and CodeBERTScore respectively. 

We find an $r_{bp}$ of $0.162$ with p-value of $3.28\times 1o^{-26}$ between CodeBERTScore and $\text{pass@1}$ and an $r_{bp}$ of $0.186$ between $\text{EDIT-SIM}$ and  $\text{pass@1}$ with a p-value of $5.83\times10^{-34}$ and an $r_{bp}$ of $0.103$ with a p-value of $2.11\times 10^{-11}$ for BERTScore.

To test \textbf{H2} we measure the auto-correlation between the results obtained with CodeBERTScore (and additionally BERTScore) before and after semantics preserving input perturbations, using Kendall Tau ($\tau$) and Spearman ($\rho$) correlations.
The perturbations are described below:
\begin{enumerate}
    \item \textbf{Var: Cand only} The variable names in the candidate are replaced with generic variable names (e.g. if \texttt{sum} is the third variable in the code, then it will be replaced with \texttt{var2}). We expect this transformation to \textbf{lower the metric value}
    \item \textbf{Var: Ref only} The variable names in the reference are replaced with generic variable names. We expect this transformation to \textbf{lower the metric value}
    \item \textbf{Var: Cand and Ref} The variable names in both reference and candidate are replaced with generic variable names. We expect this transformation to \textbf{increase the metric value}
    \item \textbf{Func: same} The function names of both candidate and reference functions are replaced with the same name (\texttt{candidate\_function}). We expect this transformation to \textbf{increase the metric value}
    \item \textbf{Func: different} The function names of both candidate and reference functions are replaced with different names (\texttt{candidate\_function} and \texttt{reference\_function} respectively). We expect this transformation to \textbf{decrease the metric value}
\end{enumerate}
We show the correlations and p-values in Table~\ref{tab:H2_results}. Note that all the p-values are $0$ and all autocorrelation values are in the strong range, even though the metric values themselves are affected slightly by the perturbations (all the trends of increasing/decreasing values are as expected).

To test \textbf{H3} we compute the Kendall Tau ($\tau$) and Spearman ($\rho$) correlations between $\text{EDIT-SIM}$ and CodeBERTScore getting $\tau=0.72$ and $\rho=0.89$ with p-values of $0$ in both cases.

\begin{table*}[]
\centering
\resizebox{\textwidth}{!}{
\begin{tabular}{@{}llrrrrr@{}}
\toprule
\multirow{2}{*}{\textbf{Metric}} &
  \multirow{2}{*}{\textbf{Transformation}} &
  \multirow{2}{*}{\textbf{F1}} &
  \multicolumn{2}{c}{\textbf{Kendall Tau Autocorrleation}} &
  \multicolumn{2}{c}{\textbf{Spearman Autocorrelation}} \\ \cmidrule(l){4-7} 
                                  &                  &                & Tau   & p-value & Rho   & p-value \\ \midrule
\multirow{6}{*}{CodeBERTScore-F1} & No transform     & 0.879          &       &         &       &         \\
                                  & Var: Cand only   & 0.847 (-0.032) & 0.811 & 0       & 0.951 & 0       \\
                                  & Var: Ref only    & 0.842 (-0.037) & 0.82  & 0       & 0.954 & 0       \\
                                  & Var: Cand \& Ref & 0.904 (+0.025) & 0.806 & 0       & 0.943 & 0       \\
                                  & Func: same       & 0.881 (+0.002) & 0.926 & 0       & 0.988 & 0       \\
                                  & Func: different  & 0.872 (-0.005) & 0.918 & 0       & 0.989 & 0       \\ \midrule
\multirow{6}{*}{BERTScore}        & No transform     & 0.937          &       &         &       &         \\
                                  & Var: Cand only   & 0.927 (-0.01)  & 0.857 & 0       & 0.97  & 0       \\
                                  & Var: Ref only    & 0.926 (-0.011) & 0.861 & 0       & 0.971 & 0       \\
                                  & Var: Cand \& Ref & 0.946 (+0.009) & 0.864 & 0       & 0.971 & 0       \\
                                  & Func: same       & 0.939 (+0.002) & 0.935 & 0       & 0.991 & 0       \\
                                  & Func: different  & 0.936 (-0.001) & 0.938 & 0       & 0.993 & 0       \\ \bottomrule
\end{tabular}
}
\caption{Results of auto-correlation under semantics preserving input perturbations for CodeBERTScore and BERTScore. The results show the although the metric values are affected slightly by the transformations, the metrics are still robust due to the high correlation between the old and new values after applying the transform.}
\label{tab:H2_results}
\end{table*}
\section{Discussion}
The results for \textbf{H1} show that while there is a statistically significant correlation between functional correctness $\text{pass@1}$ and CodeBERTScore, it is rather weak and the same is the case for BERTScore.
It is less correlated with it than the construct of editing effort $\text{EDIT-SIM}$, so we can discard \textbf{H1}.
The results for \textbf{H2} show that while the input perturbations can affect the metric values in expected ways, the metrics remain strongly and statistically significantly auto-correlated so we can accept \textbf{H2}.
The results for \textbf{H3} show a strong and statistically significant correlation so we can accept it.

From these results, we can conclude that despite claims made by \cite{zhoucodebertscore}, \textbf{CodeBERTScore (or BERTScore) is not a reliable proxy for functional correctness.} 
However, it is a relatively robust metric with the highest sensitivity (discriminative power) among reference-based metrics and could be used as a measure of syntactic similarity or editing effort for a reference/goal output.
In fact, the results indicate that embedding-based metrics might be able to capture the syntax level structure of code (suggested by the robustness to input perturbations) but fail to penetrate to the level of code semantic equivalence as supported by previous studies that probe code models \cite{troshin-chirkova-2022-probing, Naik2022ProbingSG}.
\\
\textbf{Future Work:} While this study extensively explores many state-of-the-art LLMs over two of the most popular benchmarks, it would be interesting to explore the results over more datasets like ODEX \cite{wang2022execution} that cover more challenging and open-domain code generation but still support test case evaluation to further validate the findings.
It would also be interesting to explore these results for models trained on the evaluation benchmarks for the MBPP dataset and benchmark the performance of iterative/multi-step code generators like code agents \cite{huang2024agentcoder, zhong2024ldb, hong2023metagpt, zhou2023language}.
Finally, it would be interesting to explore the effectiveness of metrics that learn execution behavior like CodeScore \cite{dong2023codescore} and explore ways to improve embedding-based metrics by learning execution behavior.
\\ \\ 
\textbf{Online Experiments:} While static benchmarks like HumanEval \cite{chen2021codex} were useful at the time of their introduction for evaluating the functional correctness of LLM-generated code, they fail to completely align with the primary use case of LLMs as programming assistants and their effect on constructs like programmer productivity.
As a result, benchmarking efforts like RealHumanEval \cite{mozannar2024realhumaneval} use an \textbf{online web interface, with an editor and chat window} to measure the ability of LLMs to assist programmers through auto-complete or chat support, while using proxy behavioral \textbf{signals like code acceptance and copy rates}.
Experiments done by \cite{mozannar2024realhumaneval} show that while increased performance on static benchmarks leads to increased productivity the gaps in benchmark and human performance are not always proportional. 
They also found that programmer preferences do not correlate with actual LLM performance motivating more human-centered evaluation.
Important \textbf{sub-populations} for such experiments include \textbf{novice programmers and experts}. 
Experts are likely to reject more code or take longer till they except a code as they might be stricter with their requirements and other aspects of code quality like clarity and efficiency.
The copy rates could be higher with any population as while experts might be more strict with accepting output they might be more likely to just copy code and prefer to edit it themselves than novice users.
Figuring out these differences for certain would require further experimentation.




\bibliography{references}
\bibliographystyle{acl_natbib}

\appendix


This is a section in the appendix.

\end{document}